\documentclass[useAMS,usenatbib]{mnras}

\usepackage{color}
\usepackage[dvipsnames]{xcolor}
\usepackage{aas_macros}
\usepackage{graphicx}
\usepackage{upgreek}
\usepackage{amssymb}

\newcommand{\msun}{{\rm M_{\sun }}}
\newcommand{\Mo}{\msun}
\newcommand{\ME}{{\rm M_\oplus}}
\newcommand{\Ro}{{\rm R_{\sun }}}

\newcommand{\sfrac}[2]{{\scriptsize \frac{#1}{#2}}}
\newcommand{\uH}{{\rm H}}
\newcommand{\vtp}{\tilde{v}_\perp}
\def\Mearth{\mathrm{M}_\oplus}

\title[An ingested super-Earth in M67 Y2235]{Super-Earth ingestion can explain the anomalously high metal abundances of M67~Y2235}
\author[R.~P.~Church, A.~J.~Mustill \& F.~Liu]{Ross~P.~Church\thanks{E-mail:
ross@astro.lu.se}, Alexander~J.~Mustill and Fan~Liu\\
Department of Astronomy and Theoretical Physics, Lund Observatory, Box 43, SE--221 00, Lund, Sweden.\\
}
\begin{document}

\date{Accepted XXX. Received XXX; in original form XXX}

\pagerange{\pageref{firstpage}--\pageref{lastpage}} \pubyear{2019}

\maketitle

\label{firstpage}


\begin{abstract}
We investigate the hypothesis that ingestion of a terrestrial or super-Earth planet could cause the anomalously high metal abundances seen in a turn-off star in the open cluster M67, when compared to other turn-off stars in the same cluster.  We show that the mass in convective envelope of the star is likely only $3.45\,\times 10^{-3}\,\Mo$, and hence $5.2\,{\rm M}_\oplus$ of rock is required to obtain the observed 0.128\,dex metal enhancement.  Rocky planets dissolve entirely in the convective envelope if they enter it with sufficiently tangential orbits: we find that the critical condition for dissolution is that the planet's radial speed must be less than 40\% of its total velocity at the stellar surface; or, equivalently, the impact parameter must be greater than about 0.9.  We model the delivery of rocky planets to the stellar surface both by planet-planet scattering in a realistic multi-planet system, and by Lidov--Kozai cycles driven by a more massive planetary or stellar companion.  In both cases almost all planets that are ingested arrive at the star on grazing orbits and hence will dissolve in the surface convection zone.  We conclude that super-Earth ingestion is a good explanation for the metal enhancement in M67~Y2235, and that a high-resolution spectroscopic survey of stellar abundances around the turn-off and main sequence of M67 has the potential to constrain the frequency of late-time dynamical instability in planetary systems.
\end{abstract}

\begin{keywords}
stars: evolution --- methods: numerical --- planet--star interactions --- planetary systems --- open clusters and associations: individual: M67
\end{keywords}

\section{Introduction}

Differential stellar spectroscopy, where the spectra of stars with very similar atmospheric parameters are analysed comparatively, is a powerful tool for testing our understanding of stellar physics and astrophysics.  It can be used to probe the evolution of stars, the physics of stellar photospheres, and the consequences of interactions between stars and their planets.  As a tool it is particularly valuable when the stars are thought to have formed from gas of the same chemical composition, and to have done so at the same time, since these additional constraints limit the possible causes of differences between the spectra.  One example is wide binary stars, where the stars can be resolved and separate spectra obtained \citep[e.g.,][]{Ramirez+11,Liu+14,Mack+14,TucciMaia+14,Ramirez+15,Teske+16a,Teske+16b}.  In a recent case \citet{Liu+18} studied the wide binary HD~80606/80607, and proposed that the difference between the two components' elemental abundances has been caused by the accretion of a planet into the envelope of HD~80606.  Suitable binaries are rare, however: they must be wide and nearby enough for the stars to be resolved, and have sufficiently similar surface temperatures to be analysed differentially.

Another source of coeval stars, whose initial chemical compositions can be assumed to be very similar, is stellar clusters.  The open cluster M67 is particularly attractive for such studies.  With over 1200 known members \citep{Geller+15}, it is rich enough to provide a useful sample of stars to study in any chosen evolutionary state.  It is also relatively nearby \citep[$\sim900$\,pc,][]{Gaia18}, making even the main sequence available to high-resolution spectroscopic analysis, and it has a metallicity very close to solar \citep[e.g.,][]{Randich+06,Pace+08,Oenehag+11,Oenehag+14,Liu+16}, allowing the Sun to be used as a reference point.

\begin{figure}
\begin{center}
\includegraphics[width=\columnwidth]{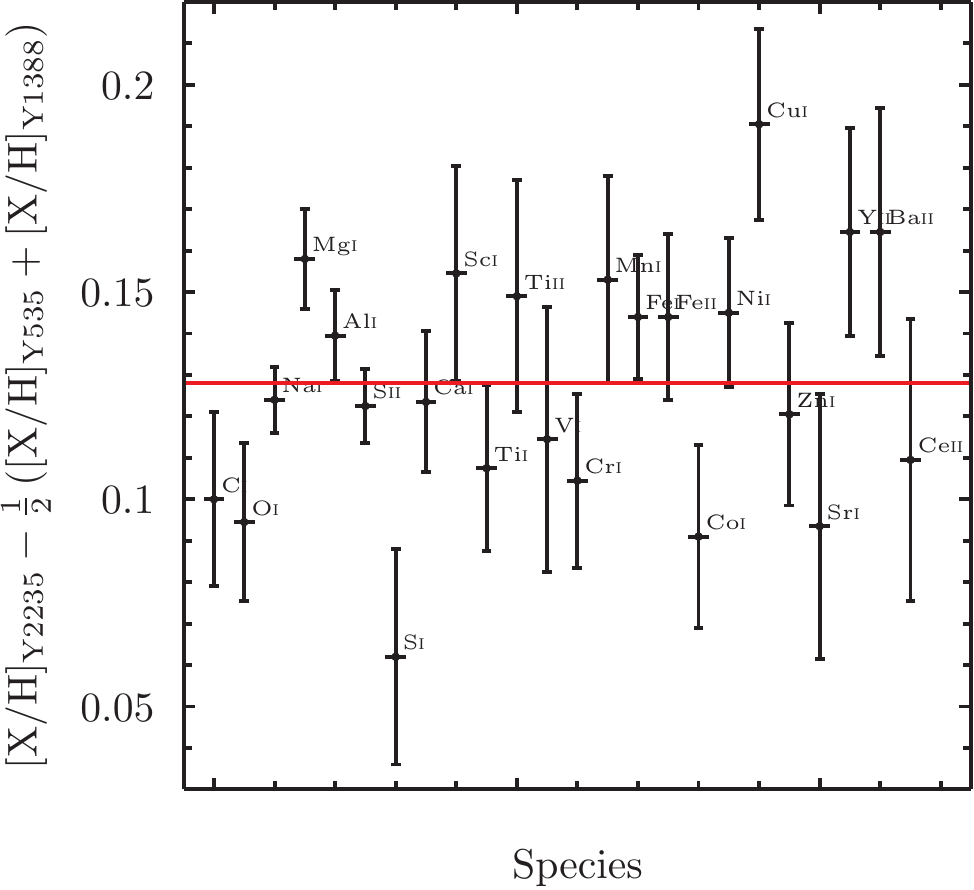}
\caption{Differences in special abundances between M67 Y2235 and the mean value for M67 Y535 and M67 Y1388.  Data are taken from tab.~3 of \citet{Liu+19}. Uncertainties are those for Y2235.  The red horizontal line shows the mean abundance offset of 0.128\,dex.}
\label{fig:abundExcess}
\end{center}
\end{figure}

\citet{Liu+19} observed seven members of M67 with high-resolution, high-signal-to-noise spectroscopy.  Their analysis is strictly differential to the Sun, which allows them to obtain relative uncertainties in elemental abundances of around 0.02\,dex.  In addition to a solar spectrum they observed one solar twin, Y1194, three turn-off stars (Y1388, Y535 and Y2235), and three subgiants (Y1844, Y519, and Y923).  We focus here on the turn-off stars.  Y1388 and Y535 have the same elemental abundances as one another to within the uncertainties ($\sim0.02$~dex), and are in qualitative agreement with the evolutionary models of \citet{Dotter+17}, which include changes to atmospheric metal abundances through atomic diffusion, radiative levitation, and turbulent diffusion.  However, the third turn-off star, Y2235, has enhanced elemental abundances compared to the other two, as shown in Figure~\ref{fig:abundExcess}.  The differences show no particular trend with atomic number or sublimation temperature \citep[unlike between some pairs of stars; see][]{Melendez+09,Melendez+17,Ramirez+15} and, to within the observational uncertainties, are consistent with a constant offset of 0.128\,dex in each species.  The presence of a systematic offset is significant to about six sigma.

Abundance differences such as these have often been attributed to the presence of extra-solar planets, in two ways: the formation of planets might lock up heavy elements in the protoplanetary disc, preventing their accretion onto the star and reducing the photospheric abundances \citep[][]{Melendez+09,Chambers10,Ramirez+15}; alternatively, the accretion of a planet might increase photospheric abundances \citep[][]{Sandquist+02,Ramirez+15,Melendez+17}. This latter process is particularly effective if the accretion takes place once the host star has reached the main sequence, as its convective envelope is then small and the material of the planet can provide a large abundance enrichment if it mixes into a relatively small fraction of the star. Often the required quantities of material are estimated to be a few Earth masses \citep[e.g.,][]{Melendez+17,Liu+18}. This places potential ingested planets in the class of super-Earths, a class which has been shown to be extremely common, with transit and radial-velocity surveys giving occurrence rates of $30-70\%$ \citep[][]{Mayor+11,Fressin+13,Zhu+18,Zink+19}. Therefore, the potential to pollute stellar photospheres by ingesting planets may exist in many systems.

Motivated by these observations, we further explore the scenario where a star's photospheric abundances are enhanced by the late accretion of a terrestrial or super-Earth planet. We focus on M67~Y2235 and investigate the evaporation and orbital decay of super-Earths that penetrate the stellar photosphere, and the initial conditions for these trajectories that arise when planets are delivered to the star by dynamical interactions with other planets or binary stellar companions. The paper is structured as follows.  In Sections~\ref{sect:StellarModels} and \ref{sect:IngestionModel} we describe our stellar models and our model for the ingestion of planets by their host star.  Section~\ref{sect:Transport} contains a discussion of the various means of transporting super-Earth planets to the stellar surface, and Section~\ref{sect:Discussion} a discussion of the consequences of our models for M67.  Finally we conclude in Section~\ref{sect:Conclusions}.

\section{Stellar models}
\label{sect:StellarModels}

\begin{figure}
\begin{center}
\includegraphics[width=\columnwidth]{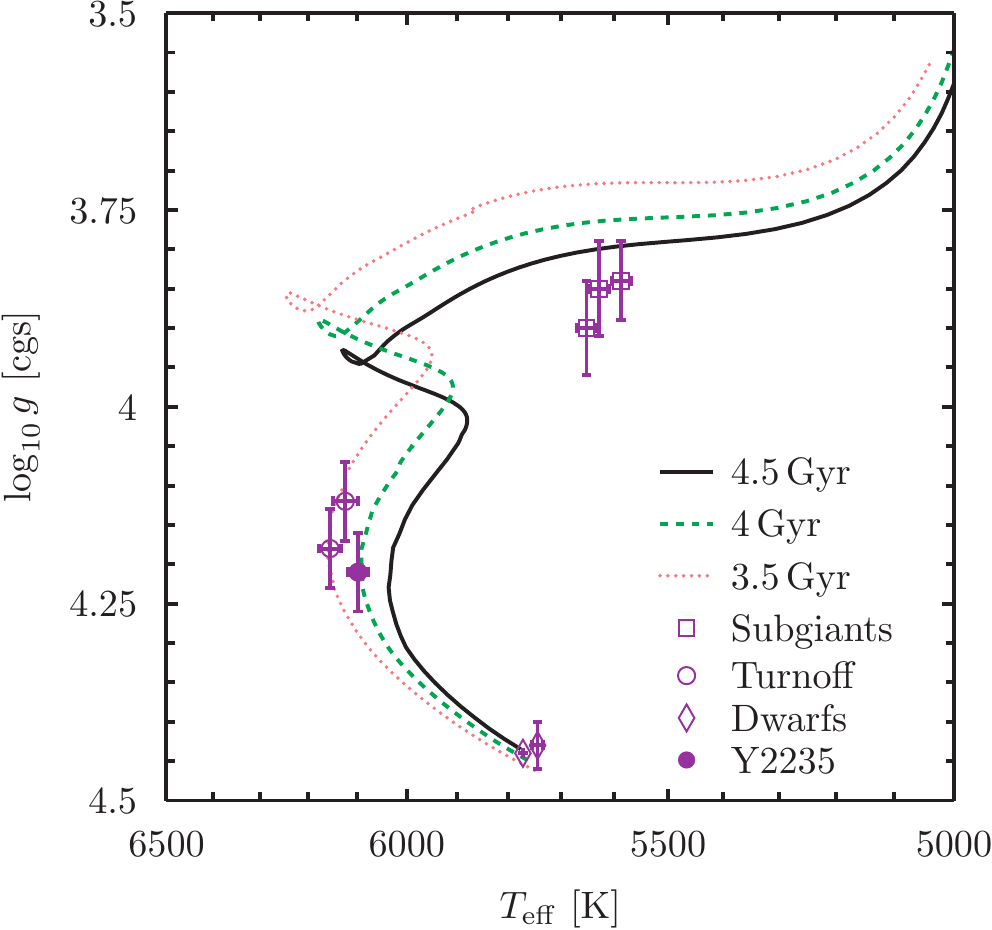}
\caption{Model isochrones generated using {\sc stars}. Surface gravity $\log_{10} g$ is plotted against effective temperature $T_{\rm eff}$.  The black solid line shows a 4.5\,Gyr isochrone, the green dashed line is for 4\,Gyr and the salmon dotted line for 3.5\,Gyr.   Purple error bars show the observations from \citet{Liu+19}: subgiants are plotted with squares, turn-off stars with circles and dwarfs with diamonds.  M67~Y2235 is shown with a filled symbol.  Isochrones are calculated for stellar masses between $1\,\Mo$ and $1.4\,\Mo$.}
\label{fig:isochrone}
\end{center}
\end{figure}

\begin{figure}
\begin{center}
\includegraphics[width=\columnwidth]{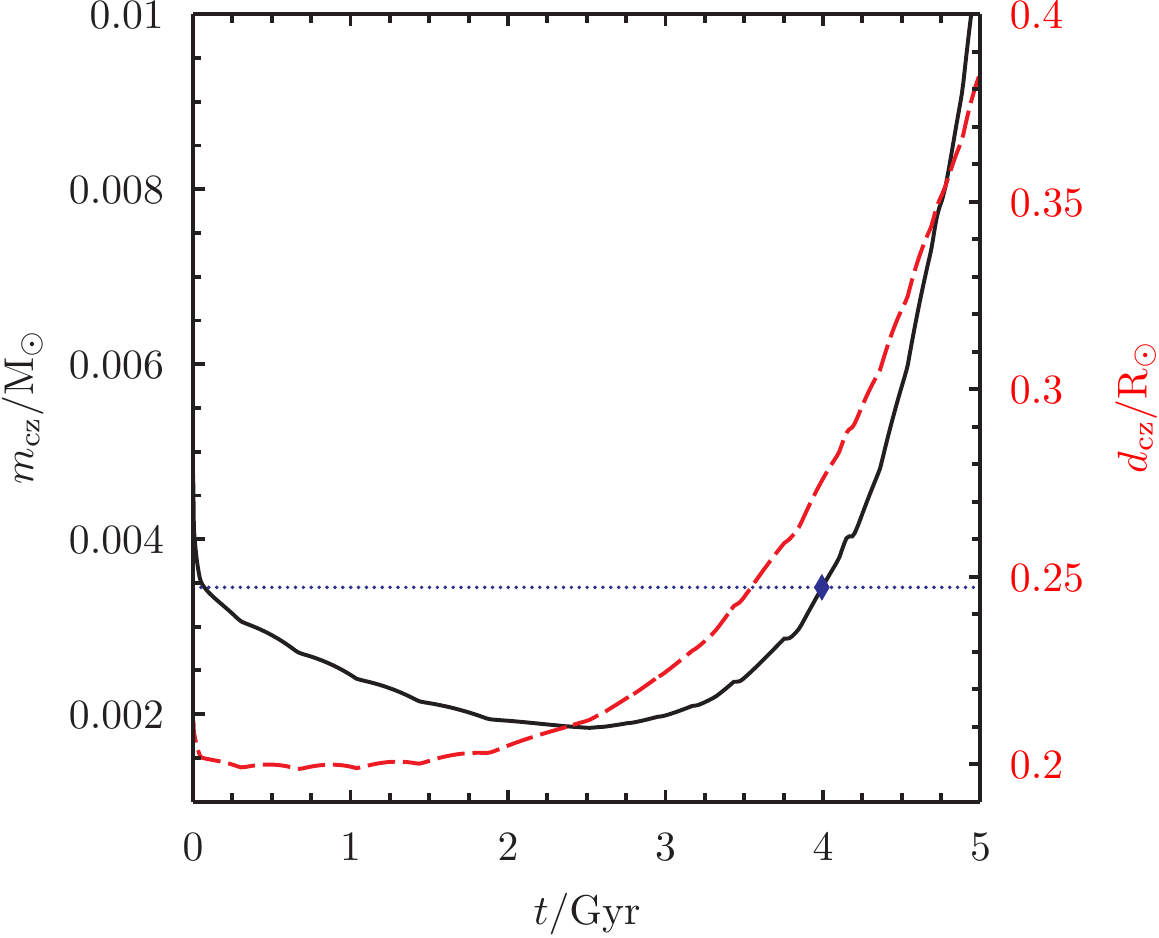}
\caption{The evolution of the convective zone of a $1.18\,\Mo$, solar-metallicity star as a function of time $t$.  The black solid line, plotted on the left-hand ordinate axis, shows the mass contained within the surface convective zone, $m_{\rm cz}$, in solar masses. The blue diamond is the turn-off model that fits M67 Y2235.  The thin dotted horizontal blue line shows the value of $m_{\rm CZ}$ at 4\,Gyr; as can be seen, other than for the brief interval whilst the star is still settling onto the zero-age main-sequence, the star's convective zone is less massive than this at all previous times. The dashed red line, plotted against the right-hand ordinate axis, shows the depth of the convective zone, $d_{\rm cz}$, in solar radii.
}
\label{fig:mCZ-1.18msun}
\end{center}
\end{figure}

We calculate stellar models using the {\sc stars} code \citep{Eggleton71,Pols+95}.  We employ standard input parameters, including solar metallicity, a convective mixing length that matches the Sun today, and convective overshooting with $\delta_{\rm os}=0.12$ \citep{Schroder+97}.  We distribute 295 models between $1\,\Mo$ and $1.4\,\Mo$ so as to resolve the features in isochrones of 3.5\,Gyr, 4\,Gyr and 4.5\,Gyr: see Figure~\ref{fig:isochrone}.  Y2235 is well fit by a 4\,Gyr old model of mass $1.18\,\Mo$, so we take that as our fiducial model for the star.  Figure~\ref{fig:mCZ-1.18msun} shows the evolution of the mass and depth of the convective envelope for this star from the zero-age main sequence to 5\,Gyr.  The star's mass is just below the point at which the stellar envelope transitions from convective to radiative, which in main-sequence stars of solar metallicity takes place at about $1.3\,\Mo$.  Hence it has a low-mass surface convective zone. 
During the first 3\,Gyr of main-sequence evolution the mass in the star's surface convective zone shrinks as the mean molecular weight of the material in the central regions increases following nuclear burning.  Towards the end of the main sequence, as the star begins to form a distinct hydrogen-exhausted core, the convective zone starts to grow again.  After 4 Gyr it deepens quickly as the star begins the transition towards the subgiant branch.  The thin blue dotted horizontal line shows that, with the exception of the first $10^8\,{\rm yr}$, the star's convective zone has never previously exceeded its mass at 4\,Gyr, which is $3.45\times 10^{-3}\,\Mo$.

The change in logarithmic number abundance of species $x$, $\Delta[x/{\rm H}]$, is, in general, given in terms of its mass fraction $X(x)$, as
\begin{equation}
\Delta[x/{\rm H}] = \log_{10}\left(\frac{X(x)X_0(\uH)}{X(\uH)X_0(x)} \right) \simeq \log_{10}\left(\frac{X(x)}{X_0(x)} \right),
\end{equation}
where $X_0$ denotes the original abundance. In practice the change in hydrogen abundance can be neglected.  If we assume that the change is caused by the accretion of $m_{\rm ac}$ of material with total metal mass fraction $Z_{\rm ac}=1$, with all the {\it individual} metals in the same ratios as the original material,\footnote{see Section~\ref{sect:Discussion} for a discussion of this assumption} then this becomes 
\begin{equation}
\Delta[x/{\rm H}] = \log_{10}\left(\frac{Z}{Z_0}\right) = \log_{10}\left(\frac{m_{\rm cz} + m_{\rm ac}/Z_0}{m_{\rm cz} + m_{\rm ac}}\right)
\end{equation}
where $m_{\rm cz}$ is once again the mass in the convective zone and $Z$ is the mass fraction of metals following accretion.  We take $Z_0=0.013$ \citep{Asplund+09} which, for our $1.18\,\Mo$ model shown in Figure~\ref{fig:mCZ-1.18msun} implies that we need to accrete a total of $1.6\times 10^{-5}\,\Mo$ of metals, about $5.2\,{\rm M_\oplus}$.

The discussion above shows that a thin convective zone permits the observed surface abundance to be changed significantly by the engulfment of a terrestrial planet. As seen in Fig~\ref{fig:mCZ-1.18msun}, the convective zone encompasses a far larger fraction of the stellar radius that it does mass: it never accounts for less than 20\% of the stellar radius.  However it is not obvious that the planet will be entirely disrupted before it sinks below the base of the convective zone.  To quantify whether the planets are disrupted in or below the convective zone, we have developed a simple model of the destruction of a planet as it sinks into the star.

\section{Disruption of ingested planets}
\label{sect:IngestionModel}

Our model of the ingestion of planets contains a number of simplifying assumptions which are discussed in Section~\ref{sect:assumptions}.  We are not attempting to make a complete model of the passage of a planet into the stellar atmosphere but to answer a simple question -- does the planet dissolve before reaching the base of the convective zone? -- for which we think that this calculation is sufficient.

\subsection{Model ingredients}
We begin our calculations at the point where the planet enters the surface of the star for the first time.  Inside the star a spherical planet of radius $R_{\rm p}$, moving with velocity $v$ relative to the star, is subject to a drag force of magnitude
\begin{equation}
F_{\rm D} = \sfrac{1}{2}C_{\rm D}\uppi R_{\rm p}^2\rho_\star(r)v^2
\end{equation}
where we set the drag coefficient $C_{\rm D}=1$ and obtain the density of the stellar surface layers $\rho_\star$ by linear interpolation in radius $r$.  We neglect the force owing to dynamical friction, $F_{\rm DF}$, since, following \citet{BinneyTremaineV2},
\begin{equation}
\frac{F_{\rm DF}}{F_{\rm D}} = \frac{4\uppi G^2 M_{\rm p}^2\rho_\star\ln\Lambda}{v^2F_{\rm D}} = \frac{2\ln\Lambda}{C_{\rm D}}\left(\frac{v_{\rm esc,p}}{v}\right)^4,
\label{eqn:DF}
\end{equation}
where the Coulomb logarithm, $\ln\Lambda \simeq 10$, and the surface escape velocity of the planet, $v_{\rm esc,p}\approx 10-20\mathrm{\,km\,s}^{-1}$, whilst $v\approx 500\mathrm{\,km\,s}^{-1}$ for a compact turn-off or main-sequence star.  We divide $F_{\rm D}$ by the mass of the planet, which we conservatively assume to have the density of iron, $\rho_{\rm p}\simeq 7.9\,{\rm g\,cm^{-3}}$, to obtain the total acceleration of the planet, $\ddot\textbf{r}$, as
\begin{equation}
\ddot\textbf{r} = -\frac{Gm_\star(r)\textbf{r}}{r^3} - \frac{3C_D}{8}\left(\frac{\rho_\star}{\rho_{\rm p}}\right)\frac{v}{R_{\rm p}}\dot\textbf{r}.
\end{equation}
The power dissipated by the drag force is equal to $F_{\rm D}v$.  Most of the liberated energy goes into heating up the stellar photosphere, but a fraction is deposited in the planet.  This energy first melts the planet and then gravitationally unbinds it; in practice the latter process requires most energy.  We represent the fraction of the energy dissipated that goes into the planet as the coefficient of radiative heat transfer, $C_{\rm H}$.  Writing the specific gravitational binding energy of the surface layers of the planet as
\begin{equation}
\epsilon_{\rm bind,p} \simeq \frac{GM_p}{R_p}
\end{equation}
we obtain the rate of loss of mass from the planet as 
\begin{equation}
\dot{M_{\rm p}} = \frac{C_{\rm H}F_{\rm D}v}{\epsilon_{\rm bind,p}+\mathcal{L}_{\rm vap}} 
\label{eqn:mdot}
\end{equation}
where we conservatively take the latent heat of vaporisation appropriate for iron, $\mathcal{L}_{\rm vap}\simeq 6\,{\rm kJ\,g^{-1}}$ \citep{zhang2011corrected}.

Obtaining a good estimate for the value of $C_{\rm H}$ is challenging.  The most relevant body of work relates to meteor motion in the Earth's atmosphere.  \citet{Brykina17} give formulae that cover the relevant regime in Mach number (roughly 50) and gas densities but for much smaller bodies.  Over their range of validity these formulae give values of $C_{\rm H}$ between roughly three and thirty per cent: hence we conservatively set $C_{\rm H}=0.01$. 

We integrate the equations for the mass and motion of the planet numerically, starting at the point at which the planet enters the star for the first time.  The total velocity of the planet at impact onto the stellar surface, $v_0$, is related to its orbital semi-major axis $a$ by
\begin{equation}
v_0=\sqrt{\frac{2GM_\star}{R_\star}\left(1-\frac{R_\star}{2a}\right)}
\end{equation}
where $R_\star$ and $M_\star$ are the stellar radius and mass.  In practice, for any realistic means of transporting the planet to the star, $a\gg R_\star$ and the planet enters the star at approximately the surface escape speed; see Section~\ref{sect:Transport} for more details.
The other initial condition that we need is the angle of incidence of the planet's orbit at the stellar surface.  We work in terms of the fraction of the planet's velocity that is in the radial direction at the surface, $\tilde{v}_\perp = v_\perp/v_0 \equiv \cos\theta$, where $\theta$ is the angle between the vector normal to the stellar surface and the velocity of the planet.  Considering the specific orbital angular momentum at the point of contact with the stellar surface, 
\begin{equation}
h = \sqrt{GM_\star a(1-e^2)} = R_\star v_0 \sin \theta
\end{equation}
we obtain $v_\perp$ as
\begin{equation}
\tilde{v}_{\perp} = \sqrt{1-\frac{a^2(1-e^2)}{R_\star(2a-R_\star)}}.
\label{eq:vtp}
\end{equation}

\begin{figure}
 \begin{center}
\includegraphics[width=\columnwidth]{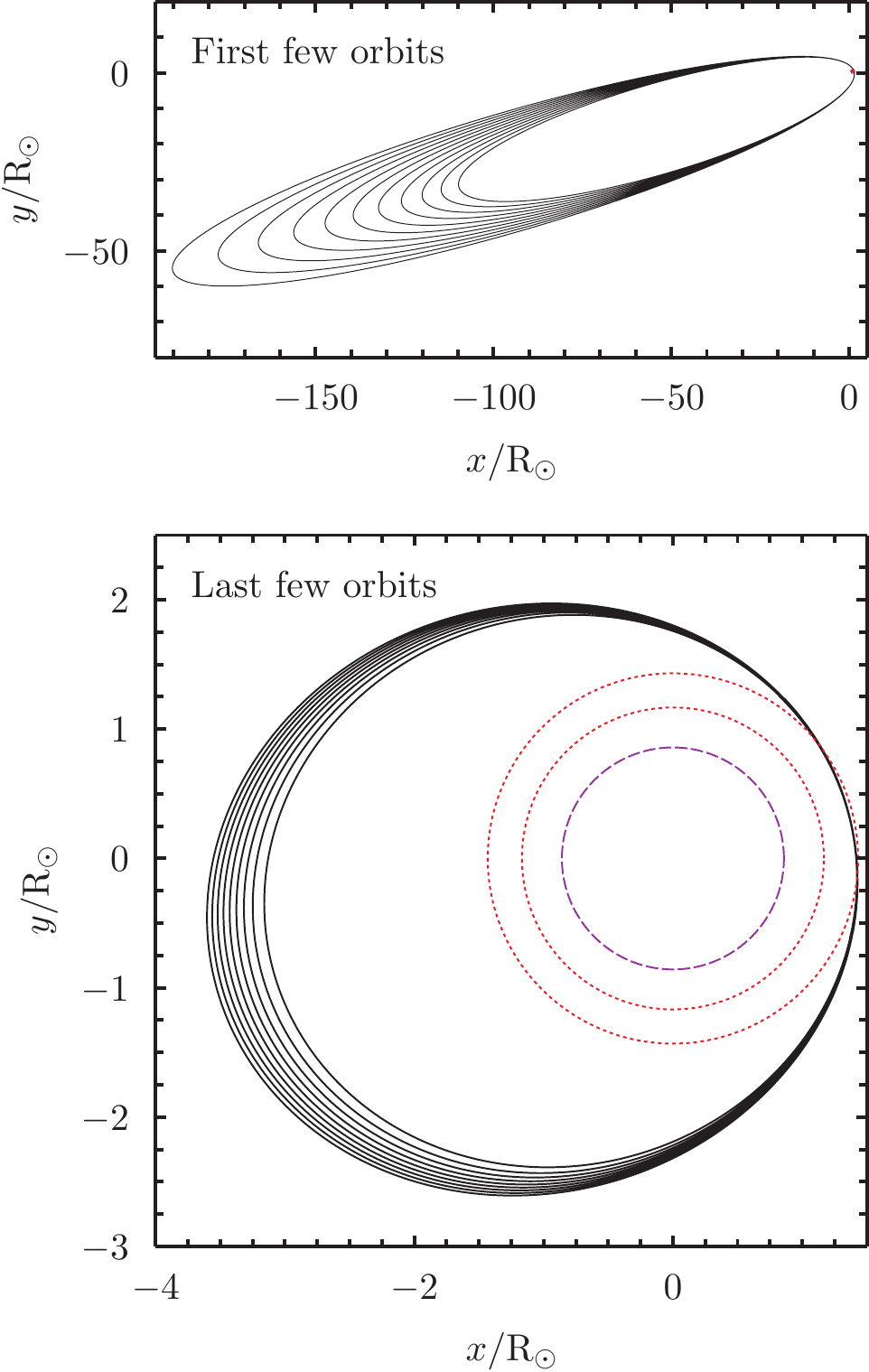}
\caption{Orbits of a planet as it dissolves in a stellar envelope.  The planet has an initial mass of ten Earth masses ($3\times 10^{-5}\,\Mo$), initial orbital semi-major axis of 0.5\,au, and 
$\vtp=0.14$, a grazing orbit.  The stellar model is a $1.18\,\Mo$, solar-metallicity star at an age of 4\,Gyr, matching the position of Y2235 in our isochrones.
The top panel shows the first few orbits after the planet enters the stellar surface; the orbital decay is caused  by the drag force as the planet passes through the envelope.  The bottom panel shows the final few orbits as the planet is disrupted.  The orbit has decayed to about twice the stellar radius (outer thin red dashed circle), but at its deepest extent it is still well above the base of the convective zone (inner thin red dashed circle). In total the planet undergoes 373 orbits before complete dissolution, taking a total of 6.4\,years.  The long-dashed purple line shows the tidal limit, within which the planet would be tidally disrupted (see Equation~\ref{eqn:Rtid}).
}
\label{fig:orbits-omb1e-2}
\end{center}
\end{figure}

\subsection{Results for fiducial model star}

Results from a typical example run are shown in Figure~\ref{fig:orbits-omb1e-2}.  We take the initial semi-major axis to be $0.5\,{\rm au}$, typical for the population of super-Earths detected by Kepler, and 
$\vtp=0.14$, which represents a grazing orbit.  The planet has an initial mass of ten Earth masses ($3\times 10^{-5}\,\Mo$). The lower panel shows the first few orbits, which are extremely eccentric: the orbit grazes the star but reaches out to a distance of roughly 1\,au from its surface.  The upper plot shows the final few orbits as the planet is disrupted.   The orbit has shrunk owing to the retarding effect of the drag force inside the envelope, but the pericentre decay is much smaller and the inner extent of the orbit is still well within the convective zone.  This planet is entirely disrupted within the convective zone and so all of its heavy metal content will be available to increase the observed metallicity of the star.

\begin{figure}
\begin{center}
\includegraphics[width=.9\columnwidth]{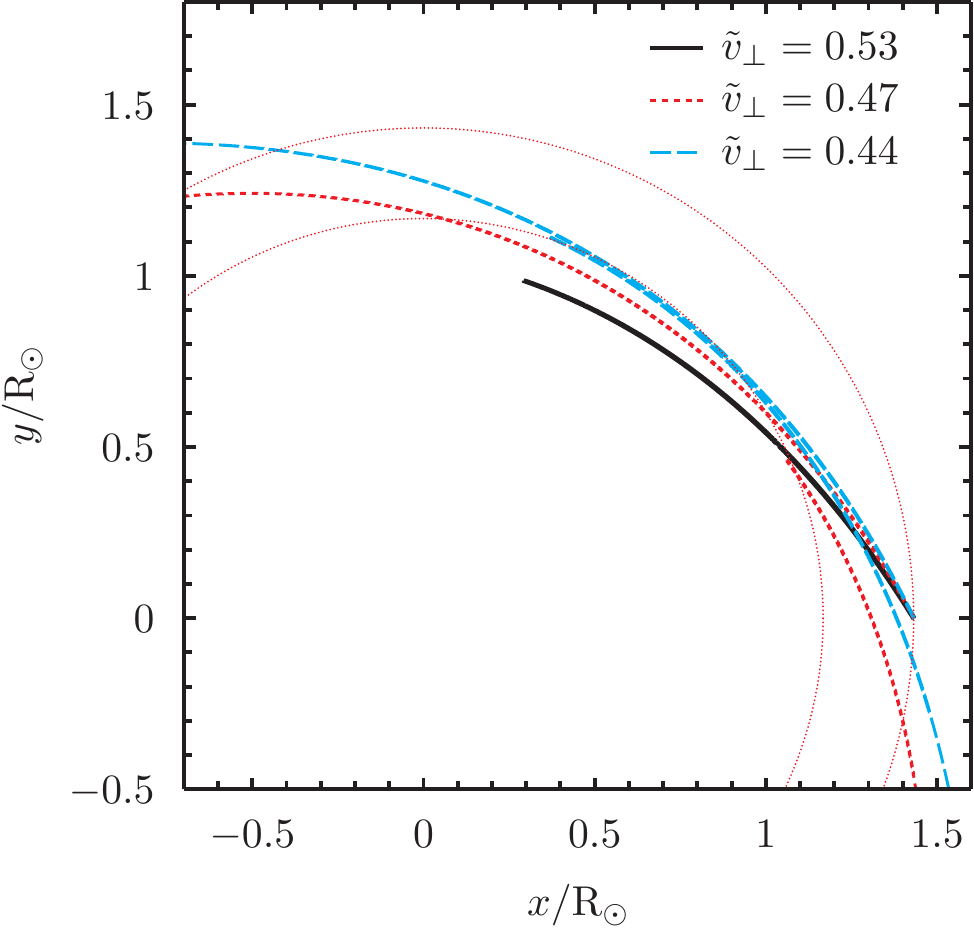}
\caption{As for Figure~\ref{fig:orbits-omb1e-2}, but for three orbits with larger values of $\vtp$.  The planet with $\vtp=0.53$ (thick, black, solid line) is destroyed on its first plunge into the star.  Planets with $\vtp=0.47$ (short-dashed red line) and $\vtp=0.44$ (long-dashed blue line) both survive the first orbit but are destroyed on the second passage.
}
\label{fig:orbits-largeb}
\end{center}
\end{figure}

By contrast, planets on more radial orbits can be destroyed rapidly.  Figure~\ref{fig:orbits-largeb} shows a few cases around the boundary between long-lived and short-lived systems for a planet mass of 10\,$M_\oplus$ ($3\times 10^{-5}\,\Mo$).  The thick black solid line shows a planet with $\vtp=0.53$; it is destroyed on its first plunge into the star and only 10 per cent of the planet's mass is deposited in the convective zone.  The case with $\vtp=0.47$ (red short-dashed line) emerges from the star but is destroyed on its second orbit; about one-third of the planet's mass ends up in the convective zone.  Finally, the blue long-dashed line shows the orbit for $\vtp=0.44$.  Once again the planet is destroyed on its second pass through the star; although the planet only grazes the radiative region below the convective zone nearly 45 per cent of its mass is deposited there. 

\begin{figure}
\begin{center}
\includegraphics[width=.9\columnwidth]{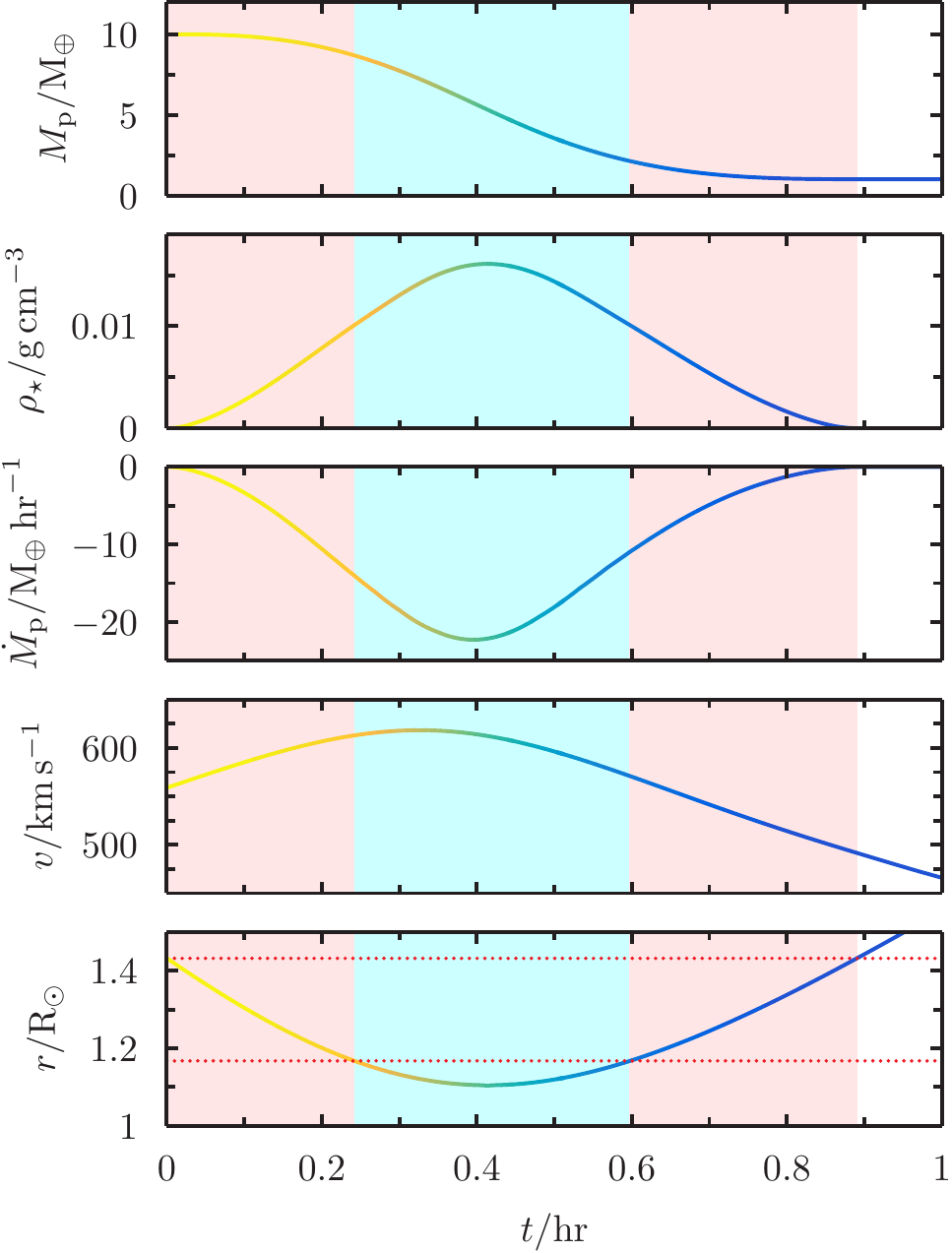}
\caption{Details of the orbit of a planet during its first pass through the star.  The initial planet mass is $10\,\ME$ and $\vtp=0.47$ (\emph{i.e.,} the short-dashed red orbit in Fig~\ref{fig:orbits-largeb}).  The bottom panel shows orbital radius $r$ as a function of time $t$, with the stellar radius and convective zone shown as dotted red lines.  Subsequent panels show the planet velocity $v$ and mass-loss rate $\dot M_{\rm p}$, the stellar density local to the planet $\rho_\star$, and the planet mass $M_{\rm p}$.  The line is coloured according to $M_{\rm p}$; the scale can be read off the top panel.  Light red shaded regions show times when the planet is inside the surface convective zone, less-light blue shaded regions when it is beneath the convective zone.
}
\label{fig:destructionDisection}
\end{center}
\end{figure}

The fact that a relatively grazing orbit can deposit most mass beneath the convective zone seems counter-intuitive, but Figure~\ref{fig:destructionDisection} shows why it is the case.  The orbit plotted is that with $\vtp=0.47$ for a $10\,\ME$ planet.  Firstly, whilst the orbit only dips below the convective zone by about $0.1\,\Ro$, it remains below the convective zone for more than a third of the total time spent inside the star since it is moving almost tangentially at that point.  Secondly, the mass loss rate $\dot M_{\rm p}\propto v^3\rho_\star$ (Equation~\ref{eqn:mdot}).  The velocity reaches its maximum whilst below the convective zone and $\rho_\star$ is a strong function of depth, making the average mass-loss rate in this part of the orbit much larger.  As a result more than two thirds of the mass deposited on the first pass ends up below the convective zone.  

\begin{figure}
\begin{center}
\includegraphics[width=\columnwidth]{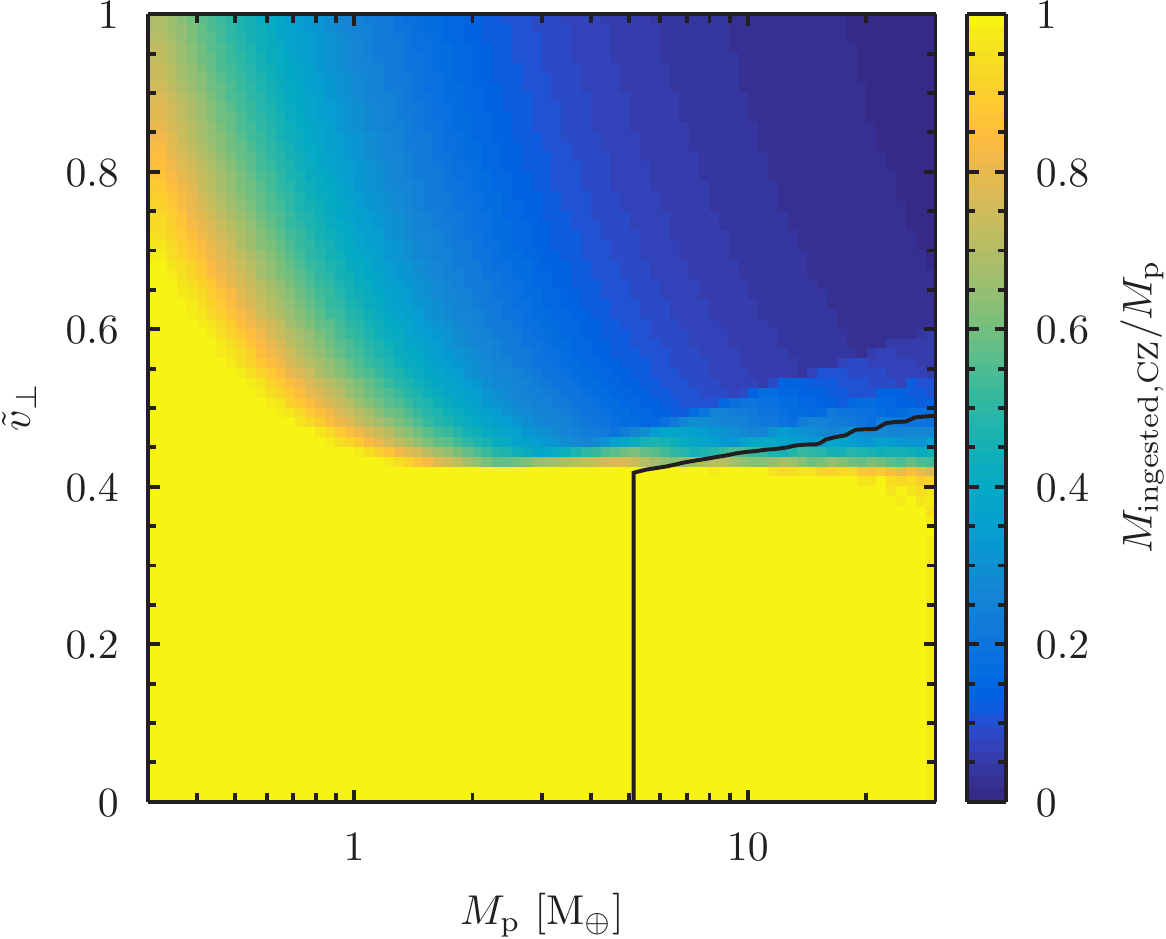}
\caption{The fraction of planetary mass dissolved in the surface convective zone of a star as a function of the initial planetary mass, $M_p$ and fractional tangential velocity $\tilde{v}_\perp$.  The black line marks the locus where enough material is deposited in the convective zone to produce the surface abundances seen in M67~Y2235.  The stellar model is a $1.18\,\Mo$, solar-metallicity star at an age of 4\,Gyr, matching the position of Y2235 in our isochrones.}
\label{fig:accretionEfficiency}
\end{center}
\end{figure}

When considering the whole parameter space of possible impacts, this effect leads to a sharp division. Planets that stay within the surface convective zone on their first orbit are, by and large, disrupted in the convective zone; hence all their heavy metal content is available to pollute the stellar photosphere.  On the other hand, planets that pass more than slightly below the convective zone are mostly disrupted in radiative regions of the star and the majority of their heavy metal content does not contribute to the photosphere.  Figure~\ref{fig:accretionEfficiency} shows this behaviour graphically. At high planet mass ($\gtrsim1\mathrm{\,M}_\oplus$) there is a relatively clean separation between planets that contribute all of their mass to the convective zone and planets where most of the mass does not end up in the convective zone.  At lower masses ($\lesssim1\mathrm{\,M}_\oplus$), the planet is mostly disrupted on the initial orbit and hence even plunging orbits can contribute most of the planet's mass to the convective zone since they are destroyed before reaching its base.  The black solid line shows the range of parameters for which enough metals are absorbed into the convective zone to change the surface abundance by 0.128\,dex, as required for M67~Y2235.

\subsection{Effects of assumptions and parameter choices}
\label{sect:assumptions}

We have made a number of assumptions in constructing this model.  
We neglect dynamical friction since Equation~\ref{eqn:DF} shows that it is insignificant compared to gas drag.  We also neglect accretional drag.  The planet is highly supersonic -- the Mach number is around 100 -- and moving through a medium much less dense than itself, so we expect accretion to be negligible.  Tidal disruption of the planet will take place when the orbit reaches approximately
\begin{equation}
a_{\rm lim,tid} = \left(\frac{3\rho_\star}{\rho_{\rm p}}\right)^{1/3}R_\star,
\label{eqn:Rtid}
\end{equation}
which for our fiducial model is about $0.86\,\Ro$, well inside the convective zone.
We also do not attempt to model the effect of the injected energy on the envelope's structure; for a calculation of likely observable effects on the star we refer the reader to \citet{Metzger+12}.

\subsubsection{Differential settling}
The analysis that we have carried out here ignores the effects of {\it differential} gravitational settling.  The results of \citet{Liu+19} show that, for most elements, gravitational settling provides a good description of the evolution of surface elemental abundances seen between the main sequence, turn-off and subgiant branch.  Our assumption here is that, since 0.128\,dex is a small abundance difference, the {\it relative} rates of settling between polluted and unpolluted stars will remain roughly the same, and hence the abundance difference will be retained through subsequent settling.  Given the uncertainties involved in the analysis this seems like a reasonable approximation to make: if it is incorrect the only likely effect is to slightly increase the mass of planet that we need to accrete.

\subsubsection{Convective overshooting}

\begin{figure}
\begin{center}
\includegraphics[width=\columnwidth]{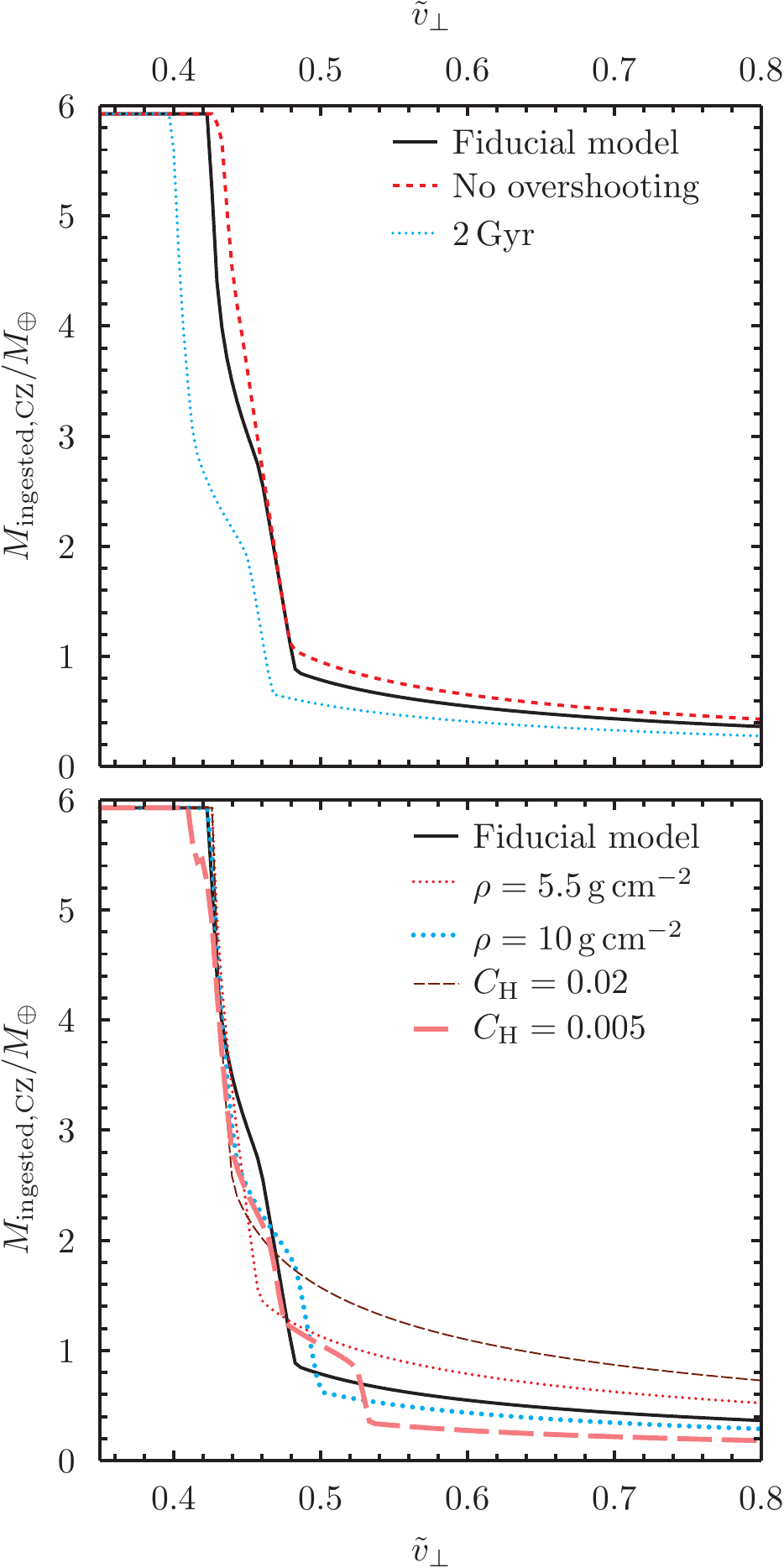}
\caption{The effects of our assumptions on the mass dissolved in the surface convective zone of a star, $M_{\rm ingested,CZ}$, as a function of the fractional tangential velocity $\tilde{v}_\perp$.  The planet has an initial mass of $6\,\Mearth$.  The stellar model is a $1.18\,\Mo$, solar-metallicity star.  The black solid lines show our fiducial model at an age of 4\,Gyr.  The {\it top panel} shows the effects of changing the properties of the star.  The red short-dashed line shows a model of the same mass calculated without convective overshooting, at an age of 3.8\,Gyr as required to match the position of Y2235.  The blue dotted line shows the fiducial star, but with an impact at 2\,Gyr rather than 4\,Gyr. The {\it bottom panel} shows the effects of changing properties related to the planet.  The planet's density is varied from the fiducial $7.9\,{\rm g\,cm^{-2}}$ to $5.5\,{\rm g\,cm^{-2}}$ (red close-dotted line) and $10\,{\rm g\,cm^{-2}}$ (blue large-dotted line).  The fraction of frictional energy liberated, $C_{\rm H}$, is varied from the fiducial 0.01 to 0.02 (brown short-dashed line) and 0.005 (pink long-dashed line).}
\label{fig:massIngestedVaryPhysicalAssumptions}
\end{center}
\end{figure}

Our models are calculated with mild convective overshooting. We set the overshooting parameter in {\sc stars} $\delta_{\rm ov}=0.12$, following the recommendations of \citet{Schroder+97} and in rough consistency with the more recent study of \citet{Stancliffe+15}.  Without overshooting we require a slightly younger model (3.8\,Gyr) to match M67~Y2235.  This is because convective overshooting increases the size of the small core convection zone which forms in this model, thereby bringing more mass into the central burning region and prolonging the life of the star.  The effect on the envelope structure when the star reaches the position of M67~Y2235 in the HR diagram, however, is sufficiently small as to make no significant difference to our conclusions.  Figure~\ref{fig:massIngestedVaryPhysicalAssumptions} shows the mass deposited into the surface convection zone as a function of $\vtp$ for a $6\,\Mearth$ planet.  The black solid line shows the fiducial model, the red dashed line the model without overshooting.  The differences are small.

\subsubsection{Time of impact}

We have modelled the impact of the planet onto M67~Y2235 as it is at its present age, but the convective envelope was shallower at previous times (see Figure~\ref{fig:mCZ-1.18msun}).  A planet transport process with a long timescale could lead to planetary engulfment at any time in the star's history.  The blue dotted line in Figure~\ref{fig:massIngestedVaryPhysicalAssumptions} shows the effect of engulfment at 2\,Gyr, when the star's accretion zone is at its shallowest extent.  Slightly less mass is deposited into the convective zone compared to the fiducial model, and the value of $\vtp$ for which all mass is deposited into the surface convection zone is reduced from 0.42 to 0.4.  As will be seen from the analysis below, this difference is small for realistic initial planetary orbits.  If some of the mass deposited between $0.02$ and $0.035\,\Mo$ from the stellar surface does not diffuse down into the interior it may be swept back up by the convective zone during its subsequent deepening.

\subsubsection{Composition of the planet}

So far we have considered a planet with a density of $\rho_{\rm p}=7.9\,{\rm g\,cm^{-2}}$, typical for iron in terrestrial conditions and approximately correct for the bulk density of a super-Earth planet.  We have also computed models for $\rho_{\rm p}=5.5\,{\rm g\,cm^{-2}}$, the density of the bulk Earth or of water-rich super-Earths \citep{Zeng+16}, and $\rho_{\rm p}=10\,{\rm g\,cm^{-2}}$, the density of cores of massive exoplanets.  These represent bounding values to the densities likely to be achieved by the terrestrial exoplanets that are relevant to our study.  In the latter case we also adopt an entropy of vaporisation of $5\,{\rm kJ\,g^{-1}}$ as appropriate for magnesium silicate \citep{Ahrens+OKeefe72}. The results are shown as the red small-dotted and blue large-dotted lines in Figure~\ref{fig:massIngestedVaryPhysicalAssumptions}.  For plunging orbits a lower-density planet deposits more mass in the convective zone, and correspondingly a higher-density planet less, and there are again differences in detail around the transition between plunging and grazing orbits; however, the basic picture where all of the mass is deposited into the surface convection zone for $\vtp<0.42$ remains the same.

\subsubsection{Radiative energy loss}

A planet on a highly eccentric orbit spends the majority of its time far from the star, and hence might be able to cool itself sufficiently to significantly reduce the rate of planetary evaporation.  The rate of change in temperature can be written in terms of the planet's luminosity $L_{\rm p}$ as
\begin{equation}
\dot{T} = -\frac{L_{\rm p}}{M_{\rm p}c_{\rm Fe}} = \frac{-3\sigma T^4}{R_{\rm p}\rho_{\rm p}c_{\rm Fe}},
\end{equation}
where $c_{\rm Fe}=0.45\,{\rm J\,g^{-1}\,K^{-1}}$ is the specific heat capacity of iron and $\sigma$ is the Stefan-Boltzmann constant.  This has a solution in terms of the initial temperature $T_0$ which takes the form 
\begin{equation}
T = T_{\rm 0}\left[1+\frac{3t}{\tau}\right]^{-1/3},
\end{equation}
where the timescale 
\begin{equation}
\tau = \frac{M_{\rm p}c_{\rm Fe}}{4\uppi R_{\rm p}^2\sigma T_0^3}\approx 250\,{\rm yr}
\end{equation}
for a $6\,{\rm M_\oplus}$ planet with $T_0=3000\,{\rm K}$.  This is comfortably longer than the orbital period which is of the order of a year or less, and hence radiative cooling can be disregarded.

\subsubsection{Summary}
These tests show that our model is robust to reasonable changes to the input physics.  Hence, if super-Earth planets can be transported to the stellar surface on sufficiently tangential orbits, their dissolution can reproduce the abundances seen in M67~Y2235.  Therefore we go on to explore briefly the processes that could lead to super-Earth transport and the parameters of the encounters that they lead to.

\section{Possible origins of the late-time ingestion of planets}
\label{sect:Transport}
In order to achieve the enrichment observed in M67~Y2235, it is necessary for the planet to be transported to the surface of the star  after the end of the pre-main sequence.  During the initial stages of its life a star possesses a giant-like structure with a deep convective envelope, driven by liberation of gravitational energy as the star contracts from its protostellar birth cloud.  For a star such as ours this pre-main-sequence phase lasts about $30\,{\rm Myr}$ \citep{Railton+14}.  If the planet enters the photosphere whilst this deep envelope is present its metal content will be mixed over a large fraction of the stellar mass and the enrichment of the surface layers will be small.  Our scenario, therefore, requires late-time dynamical activity in the planetary system of the type that we think has not occurred in the Solar System.

\subsection{Processes driving planetary delivery to the stellar surface}
\label{sect:drivingProcesses}
A likely source population providing a planetary origin for the pollution observed in the envelope of M67~Y2235 is the population of super-Earth planets on close ($\lesssim1$\,au) orbits found around a large fraction of stars. Estimates of the occurrence rate of such systems lie in the range 30--70\% \citep[e.g.,][]{Fressin+13,Zhu+18,Zink+19}, and they likely contain a range of planetary multiplicities \citep[e.g.,][]{Johansen+12,Zhu+18,Zink+19}. So the raw materials for pollution are common, requiring only a process that can bring one or more planets to the stellar surface. 

One such process is tidal decay of the orbit of an ultra-short period super-Earth such as 55~Cnc~e \citep{DawsonFabrycky10}. Using a constant $Q$ tidal model \citep{Jackson+08} with $Q^\prime_\star=10^7$, $M_\star=1.18\mathrm{\,M}_\odot$, and $R_\star=1.43\mathrm{\,R}_\odot$, we find a timescale for semimajor axis decay of $3.6$\,Gyr for a 55~Cnc~e clone ($P=0.736539$\,d and $M_\mathrm{pl}=8.08\mathrm{\,M_\oplus}$, \citealt{Demory+16}), driven by the tidal bulge raised on the star. This would have been longer in the past, as the star's radius was smaller. Tidal engulfment of ultra-short period super-Earths is also an unsatisfactory explanation because of the paucity of such planets: $\lesssim1\%$ of FGK stars host planets with periods $<1$\,d, and many are smaller than Earth \citep[e.g.][]{Hsu+18,Hsu+19}.

A more likely means of delivery is dynamical forcing by other planetary or stellar bodies in the system. We consider two broad classes of dynamical processes: Lidov--Kozai oscillations imposed by an inclined exterior companion \citep{Lidov62,Kozai62,Naoz16}, and planet--planet scattering arising from internal processes within a planetary system \citep{RasioFord96,WeidenschillingMarzari96}.

Inclined companions capable of inciting Lidov--Kozai cycles include binary stellar companions: planets have been found orbiting one component of systems at least as tight as 20 au \citep{Neuhaeuser+07}, and inclined binary companions can be acquired during dynamical encounters in the host star's birth cluster \citep{Malmberg+07}. M67~Y2235, still being a cluster member, has had a significant time to undergo such an encounter, although the rate is highest in the early years of the cluster's life.  Its radial velocity stability and spectral features imply that it is most likely single today and so any companion star must have been subsequently lost again.  Alternatively, the inclined perturber may be a planet. A planet's inclination can be increased by a close stellar encounter in the host star's birth cluster \citep{Malmberg+11,Li+19}; or it may have been captured onto a high-inclination orbit from another star \citep{Li+19}; or indeed placed there by dynamical instability in an isolated planetary system scattering a massive planet onto a high-inclination orbit.

Dynamical instability arises in systems of multiple planets with a timescale set roughly by the planets' separations \citep{Chambers+96,Marzari14} and likely driven by the overlap of three-planet mean motion resonances \citep{Quillen11}. A high rate of instability among systems of giant planets has been invoked to explain the observed distribution of orbital eccentricities \citep{JuricTremaine08,Raymond+11}, and instability in the outer system can destabilise any inner super-Earths \citep{Hansen17,Mustill+15,Mustill+17,Mustill+18}. In systems of inner super-Earths destabilised by outer giant planets undergoing planet--planet scattering, \cite{Mustill+17} found that 40\% of the loss of super-Earths was through collisions with the star. Finally, amongst systems comprising only super-Earths, a high rate of instability has been argued for to explain the small fraction of such systems in resonant chains \citep{Izidoro+17,Izidoro+19,Lambrechts+19}. 

In this paper we do not attempt to calculate the expected rate of planet engulfment by turn-off stars in M67.  Such a calculation is complex and fraught with uncertainties; furthermore, the sample to which we are comparing consists only of four stars.  Instead, having argued that engulfment should be present in some planetary systems but not ubiquitous, we model a handful of representative test cases to identify the expected orbital properties of planets that do reach the stellar envelope.

\subsection{Simulation setup}
To determine the likely velocity vectors of planets hitting stellar envelopes, we conduct $N$-body simulations using the {\sc mercury} code \citep{Chambers99}. As we need to resolve the very small pericentre passages of highly eccentric orbits, all integrations are performed using the accurate adaptive-timestep RADAU integrator with an error tolerance of $10^{-12}$, resulting in a maximum final energy error of $10^{-6}$. The stellar mass was set to $1.18\mathrm{\,M}_\odot$ and the stellar radius to $0.00666$\,au ($1.43\mathrm{\,R}_\odot$). Planets were removed upon collision with the star, at which point their position and velocity were recorded.

In our first experiment, we set up Lidov--Kozai forcing from a binary stellar companion. To accomplish this, we place a stellar companion of $0.1\mathrm{\,M}_\odot$ on an orbit with an eccentricity of $0.3$, an inclination of $80^\circ$, and a semi-major axis of 2 or 5\,au. Test particles are then distributed between $0.4$ and $0.5$\,au to mimic super-Earths. While our binary separations are extremely close for circumprimary planetary systems, we find that even a binary separation of 5\,au leads to extremely grazing orbits. Any wider binary would lead to still more grazing orbits as the forcing is weaker and hence the orbital change per orbit of the innermost body becomes smaller.

In our second experiment, we set up 100 systems likely to be unstable to planet--planet scattering. The systems each possess two $10\mathrm{\,M}_\oplus$ super-Earths and three $318\mathrm{\,M}_\oplus$ giant planets on exterior orbits. The innermost super-Earth is placed at 0.5\,au and subsequent planets are spaced by $4-6$ mutual Hill radii. Initial eccentricities are set to zero and inclinations randomly chosen to be within $5^\circ$ of the reference plane.

\begin{figure}
\begin{center}
\includegraphics[width=.95\columnwidth]{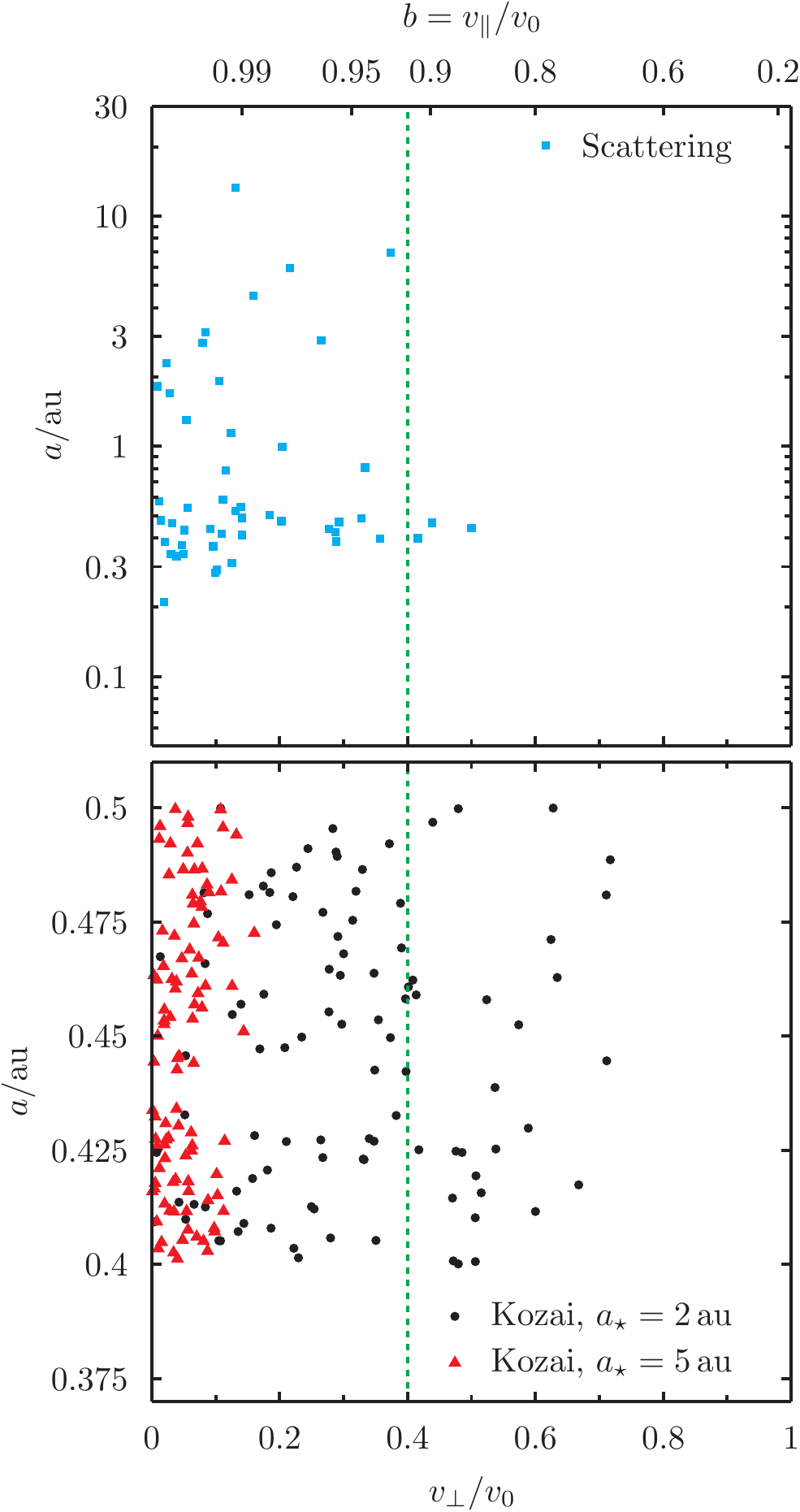}
\caption{Orbital properties of planets that impact the surface of their host stars in our $N$-body simulations.  The orbital semi-major axis at impact, $a$, is plotted as a function of the ratio of the perpendicular velocity to the total planetary velocity, $\vtp$ (lower axis), and equivalent impact parameter $b$ (upper axis).  \textit{Top panel:} scattering simulations with a system of two super-Earth planets and three Jupiter-mass planets, unstable to planet--planet scattering (blue squares).  {\it Bottom panel:} Lidov--Kozai simulations. Super-Earth planets are placed on orbits between 0.4 and $0.5\,{\rm au}$.   Lidov--Kozai cycles are forced by a binary companion of mass $0.1\,\Mo$ on orbits of 2\,au (black circles) or 5\,au (red triangles).  The green dashed vertical line shows the boundary, leftward of which all masses of terrestrial planet will be dissolved entirely within the convective envelope of our reference star.}
\label{fig:impactProperties}
\end{center}
\end{figure}

\subsection{Simulation results}

For our different simulation sets, we record the position and velocity of planets at the point when they first encounter the stellar surface, and hence derive the orbital elements and the fractional perpendicular velocity $\vtp$.  These results are shown in Figure~\ref{fig:impactProperties}. The lower panel shows simulations set up to undergo companion-driven Lidov--Kozai cycles.  The dashed green line at $\vtp=0.4$ marks the boundary where terrestrial planets of all masses are dissolved in the convective envelope of our reference model.  The simulations with a perturbing star at $2\,{\rm au}$ undergo Lidov--Kozai cycles with a short period of $\sim200-300$\,yr, meaning that the change in the planet's pericentre per orbit can be significant. As a result they have a spread of $\vtp$ values, although the median is  still as low as 0.29.  These systems represent a very extreme case unlikely to be seen in practice as exchanges into a 2-au binary in typical birth environments will be very rare.  In the cluster simulations of \cite{Malmberg+11}, who simulated relatively dense birth environments, such close binaries were never created.  On the other hand, with the perturbing companion at $5\,{\rm au}$ the cycles proceed smoothly and the planets enter the star on very tangential orbit, with a median $\vtp$ of 0.05.  In all cases the semi-major axes are not much changed from the initial values. 

The planet scattering runs produce a rather smaller number of impactors (49 from 100 systems; see top panel of Figure~\ref{fig:impactProperties}).  The planets have a much broader range of semi-major axes at impact, but the impacts are still rather grazing, with a median  $\vtp$ of 0.12.  Only three of the impactors are sufficiently eccentric to penetrate below the convective zone.  We have also run a set of simulations of scenarios similar to those proposed by \citet{Liu+18} for HD~80606/7, which yields even more grazing orbits, with a median $\vtp$ of 0.03. We find that, even in simulations where scattering takes place,  a planet is typically not delivered to the photosphere as a direct result of a strong scattering event. Rather the scattering excites the  eccentricities and inclinations of the giant planets, which in turn drive one of the super-Earth planets to the stellar surface through secular interactions. In general, the greater the change in orbital elements per orbit, the deeper can be the penetration into the stellar envelope.

\subsection{Summary}
The conclusions that we take from these simple numerical experiments are as follows.  Both Lidov--Kozai cycles induced by exchanges into binaries and by stellar flybys, and planet-planet scattering caused by inherently unstable planetary systems, can change the orbital properties of planets so that they are scattered into their host stars.  When this happens, the semi-major axes are such that the planets are travelling at very close to the stellar surface escape speed on impact.  For realistic parameters the planets dissolve in the surface convection zone of the host star, even for stars where the convective zone is a very small fraction of the total stellar mass.  Thus, in the majority of cases where terrestrial planets {\it are} scattered into a star that possesses a thin surface convection zone, pollution of the magnitude observed in M67~Y2235 should be observed.

\section{Discussion}
\label{sect:Discussion}

The analysis in the preceding sections shows that the late-time ingestion of one or more super-Earth planets is a plausible mechanism to explain the enhanced elemental abundances displayed by M67~Y2235 compared to the other two turn-off stars studied by \citet{Liu+19}.  We show that two dynamical mechanisms -- Lidov--Kozai cycles and planet--planet scattering -- can drive planets into collision with their central star.  The majority of such planets should deposit all of their mass in the surface convection zone of the star, even for a star with a thin convection zone such as M67~Y2235, so that it is available to pollute the stellar photosphere.

These processes, whilst potentially common, should not be ubiquitous.  Many stars are ``singletons'' as defined by \citet{Malmberg+07}: they have never been in a binary and have never undergone a close encounter with another star.  The Sun is one such star.  Furthermore, we know that a large fraction of the stars observed by Kepler possess dynamically cold systems of terrestrial planets, which would be disturbed by any of the processes that we discuss here.  On the other hand, as discussed in Section~\ref{sect:drivingProcesses}, there is evidence that internal dynamical instability of planetary systems may be rather common.  We do not attempt to resolve this tension here but simply note that high-resolution spectroscopic observations and differential analysis of a large sample of stars turn-off and main-sequence stars in M67 would provide a new constraint on the frequency of planet ingestion and hence dynamical instability, either internally or externally triggered, in that environment.  Other spectroscopic studies of the M67 turnoff also show potential elemental abundance variations but they are not conclusive, owing to larger uncertainties \citep{Gao+18,Souto+19}.  Whilst M67 is unusual, being a long-lived rich open cluster, such a constraint might still prove useful.  Interactions with other stars are highly unlikely to {\it stabilise} planetary systems so the rate of planet ingestion in M67 would be an upper limit of the rate of the same process in the field.  M67 is an ideal test bench owing to its age, which leads to low-mass small convective envelopes in turn-off and upper-main-sequence stars.  More massive convective envelopes would be much less sensitive to pollution and hence require unrealistically high precision in the measurements of elemental abundances.

\begin{figure}
\begin{center}
\includegraphics[width=\columnwidth]{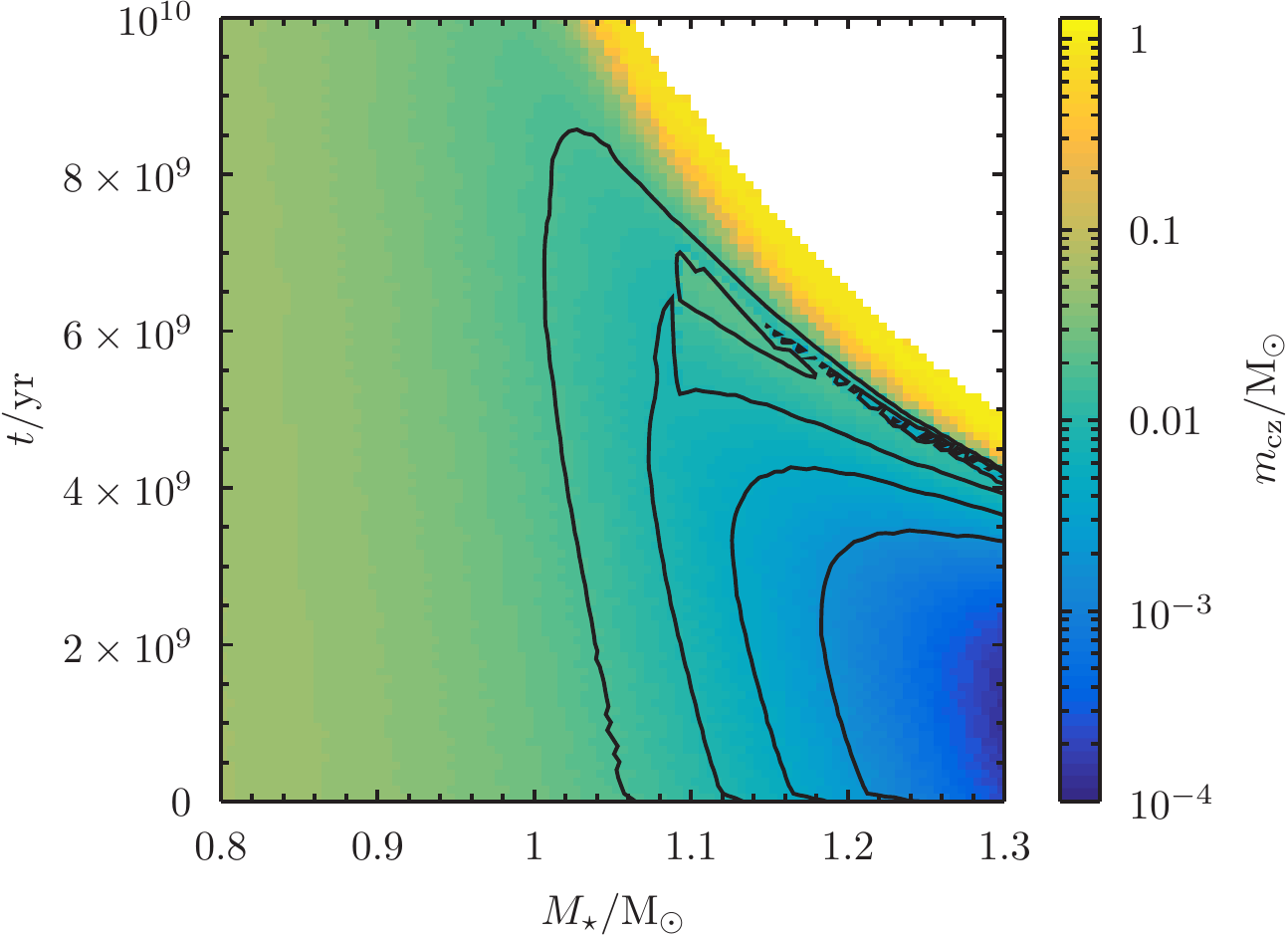}
\caption{Convective zone mass $m_{\rm cz}$ as a function of stellar mass $M_\star$ and time $t$.  The plot shows evolution to the end of the first giant branch.  The white region to the top-right of the plot is excluded by being after the end of the first giant branch.  The black lines show contours where (right to left) 2, 5, 10 and 20$\,{\rm M}_\oplus$ of polluting metals are required to change the envelope abundance by 0.1 dex.  At higher masses than shown on this figure stellar envelopes become essentially radiative on the main sequence, with only very thin convection zones driven by H and He ionisation zones.  }
\label{fig:mass-time-mcz}
\end{center}
\end{figure}

In recent years a great deal of effort has gone into the use of chemical tagging to associate stars with their birth siblings via chemical similarity \citep{Freeman_Bland-Hawthorn02}.  This study illustrates one of the limits of this technique.  Any late-time ingestion of planetary material into the envelopes of solar-like stars will lead to a spread in elemental abundances, limiting the power of the technique.  Whilst the sample size in \citet{Liu+19}, and hence this study, is small, it suggests that this is a process that should not be ignored when evaluating the potential of chemical tagging as a tool to probe the history of the Milky Way.  The problems become less pronounced when considering lower-mass stars.  Figure~\ref{fig:mass-time-mcz} shows the mass in the surface convection zones of solar-metallicity stars as a function of stellar mass and time.  Black contours enclose regions where (left to right) 20, 10, 5 and $2\,M_\oplus$ of metals are required to cause a 0.1\,dex increase in surface metal abundance.  This shows that for stars of solar mass and below, elemental abundance changes owing to accretion of rocky planets are likely to be dwarfed by uncertainties owing to our incomplete understanding of atomic diffusion, radiative levitation, and convective motions below the base of the surface convection zone.

A similar caveat applies to the use of stellar abundances as a prior for narrowing down planetary structure and composition models consistent with an observed mass and radius \citep[e.g.,][]{Dorn+17}. The enrichment of M67~Y2235, around $0.1$~dex, can be comparable to the uncertainty in abundances of exoplanet host stars. Ingestion of a planet whose abundances did \emph{not} match those of the host star will bias the inferences about the compositions and interiors of any surviving observed planets. The effect is relatively minor on bulk properties such as envelope and water fractions, but can be significant for properties of planetological significance such as core radii and the iron fraction of the mantle \citep[see][fig~8]{Dorn+17}. On the other hand, systems which have likely \emph{not} undergone a dynamical instability, such as TRAPPIST-1 \citep[analysed by][]{Dorn+18}, should be unaffected by such concerns.

Nonetheless, the abundance enrichment of M67~Y2235 is, to the precision available, the same for all species measured (see Fig~\ref{fig:abundExcess}), implying the ingestion of a planet of broadly stellar composition (albeit depleted in H/He). In particular, and unlike the Earth, C and O are not significantly depleted below the stellar abundance. This likely represents the ingestion of an icy/water-rich super-Earth, whose orbit prior to ingestion could have been outside the ice line close to where it formed, or inside the ice line if it had migrated extensively in the protoplanetary disc. Current work on the photoevaporative loss of atmospheres of planets with orbital periods of up to a few tens of days favours a rocky, not icy, core composition for these planets \citep{OwenWu17}. The planet ingested by M67~Y2235 therefore likely originated on a wider orbit.

\section{Conclusions}
\label{sect:Conclusions}

M67~Y2235 is a turn-off star in M67 which has elemental abundances enhanced by an average of 0.128\,dex compared to other turn-off stars in the cluster, and compared to models of atomic diffusion and levitation which fit other stars on the cluster main sequence.  The enhancement shows no trend with elemental mass or condensation temperature.  We have investigated whether these enhanced abundances can be reproduced by the ingestion of planets into the surface layers of the star.  Ingestion of a total of $5.2\,M_\oplus$ of metals is required.  We have constructed a simple model of the dissolution of a terrestrial planet in the outer layers of such a star.  We find that terrestrial and super-Earth planets, with masses less than around $30\,M_\oplus$, should be dissolved entirely in the surface convection zone of the star, as long as their orbits are sufficiently grazing that $\vtp=v_\perp/v \lesssim 0.4$ when the planet enters the stellar surface for the first time.  We further investigate the orbital properties of planets delivered to the stellar surface by Lidov--Kozai cycles and by scattering arising in internal dynamical instabilities within the planetary system, and show that for realistic systems these processes produce orbits that are overwhelmingly sufficiently grazing to have $\vtp<0.4$.  Hence the planets so delivered should contribute their metals to pollute the photosphere, even when the surface convection zone is as low-mass as in M67~Y2235.  As such, ingestion of planets can explain the elemental enrichment seen in M67~Y2235.  We suggest that high-resolution, high-SNR spectroscopy of a large sample of stars around the turn-off and upper-main sequence of M67 would provide a valuable dataset to investigate the frequency of planet ingestion owing to dynamical instability

\section*{Acknowledgements}
The authors would like to thank Melvyn B.~Davies, Michiel Lambrechts, and Hans Zinnecker for helpful comments.
RPC and AJM are supported by the Swedish Research Council (grants 2017-04217 and 2017-04945), and by the project grant 2014.0017 ``IMPACT'' from the Knut \& Alice Wallenberg Foundation. Fan Liu is supported by the grant ``The New Milky Way'' from the Knut \& Alice Wallenberg Foundation and grant 184/14 from the Swedish Research Council. Calculations presented in this paper were carried out using resources provided by the Swedish National Infrastructure for Computing (SNIC) at Lunarc, using computer hardware funded in part by the Royal Fysiographic Society of Lund.

\bibliographystyle{mnras.bst}
\bibliography{new.bib}

\label{lastpage}

\end{document}